\documentclass{PoS}
\PoS{PoS(HEP2005)182}
\title{Physics Reach with a Monochromatic Neutrino Beam from Electron Capture}

\ShortTitle{ Monochromatic Neutrino Beams}

\author{J.~Bernabeu,~ J. ~Burguet-Castell,~ \speaker{C. ~Espinoza}\\

       Universitat de Val\`encia and IFIC, E-46100 Burjassot, Val\`encia, Spain\\

       E-mail: \email{Jose.Bernabeu@uv.es},\\
               \hspace{1.1cm} \email{jordi.burguet.castell@cern.ch},\\
               \hspace{1.1cm} \email{m.catalina.espinoza@uv.es}}

\author{M.~Lindroos\\

       AB-division, CERN, Geneva, Switzerland\\

       E-mail: \email{Mats.Lindroos@cern.ch}}

\abstract{Neutrino oscillation experiments from different sources have
demonstrated non-vanishing neutrino masses and flavour mixings. The next
experiments have to address the determination of the connecting mixing U(e3) and the
existence of the CP violating phase. Whereas U(e3)
measures the strength of the oscillation probability in appearance
experiments, the CP phase acts as a phase-shift in the interference pattern. Here we
propose to separate these two parameters by energy dependence, using the novel idea of
a monochromatic neutrino beam facility based on the acceleration of ions that
decay fast through electron capture. Fine tuning of the boosted neutrino energy allows
precision
measurements able to open a window for the discovery of CP violation, even for
a mixing as small as 1 degree.}

\FullConference{International Europhysics Conference on High Energy Physics\\

		 July 21st - 27th 2005\\

		 Lisboa, Portugal}

\begin{document}

\section{Introduction}

\noindent Spectacular results have been obtained in the last few years for the flavour mixing of neutrinos obtained from atmospheric, solar, reactor and accelerator sources. Next experiments able to measure the still undetermined mixing and the $CP$ violating phase $\delta$ need to enter into a high precision era with new machine facilities and very massive detectors. The best way to measure the mixing parameters and $\delta$ is through appearance experiments. Signals of $CP$ violation are specially enhanced in the subleading transitions. The appearance probability $P(\nu_e \to \nu_{\mu})$ as a function of the distance between source and detector $(L)$ is given by \cite{cervera}
\begin{eqnarray}\label{prob}
P({\nu_e \rightarrow \nu_\mu}) & \simeq &
s_{23}^2 \, \sin^2 2 \theta_{13} \, \sin^2 \left ( \frac{\Delta m^2_{13} \, L}{4E} \right ) +
c_{23}^2 \, \sin^2 2 \theta_{12} \, \sin^2 \left( \frac{ \Delta m^2_{12} \, L}{4E} \right )
\nonumber \\
& + & \tilde J \, \cos \left ( \delta - \frac{ \Delta m^2_{13} \, L}{4E} \right ) \;
\frac{ \Delta m^2_{12} \, L}{4E} \sin \left ( \frac{  \Delta m^2_{13} \, L}{4E} \right ) \, ,
\end{eqnarray}
where $\tilde J \equiv c_{13} \, \sin 2 \theta_{12} \sin 2 \theta_{23} \sin 2 \theta_{13}$,
$s_{ij}$ and $c_{ij}$  are the corresponding $sin$ and $cos$ functions of $\theta_{ij}$ and
$\Delta m_{ij}^2$ the square mass differences. The four measured parameters
 $(\Delta m_{12}^2,\theta_{12})$  and  $(\Delta m_{23}^2,\theta_{23})$ have been fixed throughout this work to their mean values \cite{concha}.

Neutrino oscillation phenomena are energy dependent for a fixed distance between source and detector (see Eq. \ref{prob}),
and the observation of this energy dependence would disentangle the two important parameters:
whereas $\vert U_{e3} \vert$ gives the strength of the appearance probability, the $CP$ phase
acts as a phase-shift in the interference pattern.
These properties suggest the consideration
of a facility able to study the detailed energy dependence by means of fine tuning of monochromatic
neutrino beams. Electron-capture from short-lived isotopes has been proposed \cite{bernabeu}. In such a facility, the neutrino energy is dictated by the chosen boost
of the ion source and the neutrino beam luminosity is concentrated at a single
known energy. Such beams will concentrate all the intensity in a single energy, which may be chosen at
will for the values in which the sensitivity for the $(\theta_{13}, \delta)$ parameters is higher. Here we propose a possible implementation of the concept which would involve the use of EURISOL to produce the unstable ions, the SPS to accelerate them, and a decay ring, much like the one proposed for $\beta$-beams \cite{zucchelli}.

\vspace{-0.1cm}

\section{Neutrinos from electron capture}

\noindent Electron Capture is the process in which an atomic electron is captured by a proton
of the nucleus leading to a nuclear state of the same mass number $A$, replacing the
proton by a neutron, and a neutrino. Kinematically, it is a two body decay and so the energy of the resulting neutrino is fixed. Thus, from the single energy $e^{-}$-capture neutrino spectrum, we can get a pure and monochromatic beam by accelerating ec-unstable ions. This would be feasible if we had ions that decay fast enough to allow electron capture to occur. Recent discovery of nuclei far from the stability line having super allowed spin isospin transitions to Gamow-Teller resonances turn out to be very good candidates \cite{algora}. In Table \ref{tab:1}  we show the properties of a few ion candidates.

\begin{table}[ht!]
\centerline{
 \begin{tabular}{|l|l|l|l|l|l|l|l|}
  \hline\hline
  Decay & $T_{1/2}$ & $BR_{\nu}$ & $EC/{\beta^+}$ & $E_{\nu}$ (KeV) \\ \hline
$^{148}Dy \rightarrow ^{148}Tb^{*}$ & $3.1 m$ & $1$ & $96/4$ & $2062$ \\ \hline
$^{150}Dy \rightarrow ^{150}Tb^{*}$ & $7.2 m$ & $0.64$ & $100/0$ &  $1397$ \\ \hline
$^{152}Tm2^{-} \rightarrow ^{152}Er^{*}$ & $8.0 s$ & $1$ & $45/55$ &  $4400$  \\ \hline
$^{150}Ho2^{-} \rightarrow ^{150}Dy^{*}$ & $72 s$ & $1$ & $77/33$ & $3000$  \\ \hline
 \end{tabular}}
\caption{Four fast decays in the rare-earth region above $^{146}Gd$ leading to the
 giant Gamow-Teller resonance. $T_{1/2}$ is the life-time, $BR_{\nu}$ the branching ratio of the decay to neutrinos, $EC/{\beta^+}$ the relative branching between EC and $\beta^+$ and $E_{\nu}$ is the neutrino energy. }
\label{tab:1}
\end{table}

\vspace{-0.1cm}

\section{Neutrino Flux}

\noindent A neutrino of energy $E_0$ that emerges from radioactive decay in an accelerator will
 be boosted in energy. The measured  energy distribution can
be expressed as $E = E_0 / [\gamma(1- \beta \cos{\theta})]$ where the angle $\theta$ is the deviation between the actual neutrino detection and the ideal detector position. The neutrinos are concentrated inside a narrow cone around the forward direction. The energy distribution of the Neutrino Flux arriving to the detector in absence of neutrino oscillations is given by the Master Formula
\begin{equation}\label{master}
\frac{d^2N_\nu}{ dS dE}
= \frac{1}{\Gamma} \frac{d^2\Gamma_\nu}{dS dE} N_{ions}
\simeq \frac{\Gamma_\nu}{\Gamma} \frac{ N_{ions}}{\pi L^2} \gamma^2
\delta{\left(E - 2 \gamma E_0 \right)},
\end{equation}
with a dilation factor $\gamma >> 1$. In the last equation, $N_{ions}$ is the total number of ions decaying to neutrinos and $\Gamma_{\nu}/\Gamma$ is the branching ratio for electron capture.
In the forward
direction the neutrino energy is fixed by the boost $E = 2 \gamma E_0$, with the entire neutrino
flux concentrated at this energy. The number of events will increase with higher
 neutrino energies as the cross section increases with energy.

 A facility incorporating such technique will measure the neutrino
oscillation parameters by changing the $\gamma$'s of the decay ring (energy dependent measurement) so that there is no need of energy reconstruction in the detector.

\section{Physics Reach}

\noindent We have made a simulation study in order to reach conclusions about the measurability of the unknown oscillation parameters. The ion type chosen is $^{150}Dy$, with neutrino energy at
rest given by 1.4 MeV due to a unique nuclear transition from $100 \%$ electron capture in
going to neutrinos. For our simulation we have used a source of $10^{18}$ ions/year, during a total running time of 10 years. We used $5$ years at an ion acceleration of $\gamma_{max} = 195$, the maximum energy achievable at CERN's SPS, and  $5$ years running at $\gamma_{min} = 90$, in order to avoid background in the detector below a certain energy. The detector has an active mass of $440$ $kton$ and is located at a distance $L = 130$ $km$ (CERN-Frejus). Statistics from both appearance and disappearance events is accumulated.

\begin{figure}
\centering
\includegraphics[width=.6\textwidth]{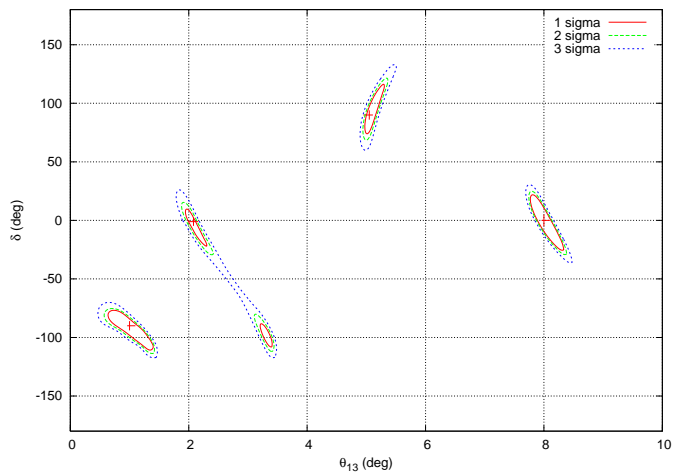}
\caption{Physics Reach for the presently unknown $(\theta_{13},\delta)$ parameters,
 using two definite energies in the electron-capture facility discussed in this paper.}
\label{fits}
\end{figure}

The Physics Reach is represented by means of the plot in the parameters $(\theta_{13}, \delta)$ as given in Fig. \ref{fits}, with the expected results shown as confidence level lines for the assumed values $(8^{\circ}, 0^{\circ})$, $(5^{\circ}, 90^{\circ})$, $(2^{\circ}, 0^{\circ})$ and $(1^{\circ}, -90^{\circ})$.

 The improvement over the standard beta-beam reach is due
to the judicious choice of the energies to which the intensity is concentrated:
whereas $\gamma=195$ leads to an energy above the oscillation peak with almost no dependence of
the $\delta-$phase, the value $\gamma=90$, leading to energies between the peak and the node,
is highly sensitive to the phase of the interference. These two energies are thus complementary
to fix the values of $(\theta_{13},\delta)$.

\section{Conclusions and Prospects}

\noindent The discovery of isotopes with half-lives of a few
minutes or less, which decay mainly through electron-capture to Gamow-Teller resonances, certainly opens the possibility for a monochromatic neutrino
 beam facility which is well worth exploring. The capacity of such a facility to discover new physics is impressive, allowing precision measurements even for a $[U_{e3}]$ mixing as small as $1$ degree. Therefore, the principle of an energy dependent
 measurement is working and a window is open to the discovery of CP violation in
 neutrino oscillations. We are aware that the concept based on monochromatic neutrino beams needs further exploration, both in machine aspects and in physics conditions. For instance, a detailed study of production cross-sections and the approach to acceleration and storage of the ions is required in order to reach a definite answer on the achievable flux.
 On the other hand, preliminary results presented elsewhere show that a combination with a $\beta^{-}$-beam is very promising for good measurements of $\delta$ \cite{jordi}. Certainly, our results encourage the further development of the concept of monochromatic neutrino beams.

\section*{Acknowledgements}

\noindent This research has been funded by the Grants FPA2004-20058E, GV05/264 and FPA2002-00612.

\end{document}